\documentclass[12pt]{article}
\usepackage[a4paper,bindingoffset=0.2in,%
            left=1in,right=1in,top=1in,bottom=1in,%
            footskip=.35in]{geometry}
\usepackage{setspace}
\usepackage{graphicx}
\usepackage{authblk}                 

\usepackage{float} 
\usepackage[caption = false]{subfig}
\usepackage{color}

\begin{document}

\title{
Modelling the influence of progressive social awareness, lockdown and anthropogenic migration on the dynamics of an epidemic
} 
\author[]{R. Bhattacharyya\thanks{ramit@prl.res.in} }
\author[]{Partha Konar\thanks{konar@prl.res.in} }
\affil[]{Physical Research Laboratory, Ahmedabad - 380009, Gujarat, India}
\date{}                     

\maketitle

\begin{abstract}
The basic Susceptible-Infected-Recovered (SIR) model is extended to include effects of progressive social awareness, lockdowns and anthropogenic migration. 
It is found that social awareness can effectively contain the spread by lowering the basic reproduction rate  $R_0$. Interestingly, the awareness is found to be more effective in a society which can adopt the awareness faster compared to the one having a slower response. The paper also separates the mortality fraction from the clinically recovered fraction and attempts to model the outcome of lockdowns, in absence and presence of social awareness. It is seen that staggered exits from lockdowns are not only economically beneficial but also helps to curb the infection spread. Moreover, a staggered exit strategy with progressive social awareness is found to be the most efficient intervention.
The paper also explores the effects of anthropogenic migration on the dynamics of the epidemic in a two-zone scenario.
The calculations yield dissimilar evolution of different fractions in different zones. 
Such models can be convenient to strategize the division of a large zone into smaller sub-zones for a disproportionate imposition
of lockdown, or, an exit from one. Calculations are done with parameters consistent with the SARS-COV-2 pathogen in the Indian context.
\end{abstract}


\newpage
\section{Introduction}
\label{sec:into}
The mathematical modelling of infectious disease is necessary to understand its spread 
among a population as the individuals interact among themselves. Additional to various 
transmission mechanisms and properties of the pathogen, the spread can also be
a function of societal properties which can include social habits, travel patterns, social distancing and 
personal hygiene. The models---stand-alone 
or combined with statistical techniques---provide insights related to the severity
of infection by predicting the number of 
infected persons, the rate at which they are getting infected and the mortality rate; among others. The 
information can further be employed to strategize various interventions in advance to contain the spread. For 
example, in 
the ongoing COVID19 pandemic \cite{covid19} in India, interventions in the form of 
early screenings and isolations along with the ultimate lockdown---claimed by WHO to 
be "timely and toughest" \cite{lancet}--- are implemented.

Effective, but mathematically straightforward, are the compartmental models which assign individuals of a population at a particular stage of the epidemic to designated compartments \cite{compmod}. 
Governed by ordinary differential equations (ODEs), individuals are then allowed to move from one compartment to another as they pass through various stages of the epidemic. The number of compartments, their coupling and the inter-compartmental flow is decided by various properties of the concerned pathogen; including its incubation period and the duration of immunity in the recovered patients--- along with other external factors like the
availability of a vaccine or the number of births and deaths during the evolution. The models inherently assume individuals in a particular compartment to be characteristically identical. Such an assumption is possible only when the population is large enough to make the probability of distributing identical individuals in a compartment statistically significant. Consequently, the compartmental models are expected to work well for systems having large populations. The simplest of the compartmental models are the SIR model,
first used by Kermack and McKendrick in 1927 \cite{sir27} and subsequently applied to a variety of diseases, 
especially airborne childhood diseases with lifelong immunity upon recovery---like measles, mumps, 
rubella and pertussis (see \cite{roz} and references therein). In its basic form, the model lacks vital dynamics, {\it i.e.} does not take into account the births/deaths
along with the incubation period of the pathogen and the recurrence of susceptibility in completely recovered individuals. Further extensions of this model (Susceptible - Exposed - Infectious - Recovered (SEIR)  
and Susceptible - Exposed - Infectious - Recovered - Susceptible  (SEIRS) are made to include the long incubation periods of certain pathogens ( like chickenpox and dengue) during which an individual can be infected but not infectious. A comprehensive list of these models along with their governing ODEs can be found in \cite{web:idmod_model}, 
the hosting site of the Epidemiological MODeling software (EMOD)---developed and maintained by 
The Institute for Disease Modeling (IDM), an institute within the Global Good Fund---a collaboration between 
Intellectual Ventures and Bill and Melinda Gates; idem \cite{web:idmod}.
Although the models described above are highly sophisticated, but we believe extensions are required to integrate societal and behavioural
changes in response to an epidemic. Toward the objective, 
we consider the SIR model as the baseline for its mathematical simplicity. Additionally, we also add inter-
zone dynamics due to migration and a procedure to calculate mortality among the recovered individuals. 
The organization of the paper is as follows: Section~\ref{sec:model} introduces Initial
Value Problems (IVPs) with their governing ODEs, Section~\ref{sec:simulation} documents the simulations
and analyze the results while Section~\ref{sec:summary} summarizes the important findings. 

\section{The Initial Value Problem}
\label{sec:model}
 
Since the proposed IVPs are based on the SIR model, in the following, we introduce the model ODEs to lay the basis for their attempted advancements \cite{sir27}. With $N$ as the total population, 
variables $S$, $I$ and $R$ denote the number of individuals who are Susceptible (not infected),
Infected and Recovered at an instant $t$. The corresponding fractions are

\begin{eqnarray}
\label{sir}
& &s(t)=\frac{S}{N},\\
& &i(t)=\frac{I}{N},\\
& &r(t)=\frac{R}{N},
\end{eqnarray}
which can also be interpreted as probabilities satisfying 

\begin{equation}
\label{conserv}
s+i+r=1,
\end{equation}
in the absence of any external forcing, {\it i.e.} no change in population because of birth/death or
migration. The rate of change of $S$ is directly proportional to the fraction $s(t)$ and the total number of infected $I(t)$, yielding. 

\begin{equation}
\label{rateS}
\frac{dS}{dt} = -b \, s(t) \, I(t) ~,
\end{equation}
where b is the proportionality constant. 
Realizing, the infected ultimately get recovered 
(or removed, because of death) 

\begin{equation}
\label{rateR}
\frac{dR}{dt} = -k \, I(t) ~,
\end{equation}
provided the recovered individuals acquire a permanent immunity to the pathogen
and, there is no delay between the exposure and getting infected. Dividing both
sides with $N$, a convenient form is 

\begin{equation}
\label{rates}
\frac{ds}{dt} = -b \, s(t) \, i(t) ~,
\end{equation}

\begin{equation}
\label{rater}
\frac{dr}{dt} = k \, i(t) ~.
\end{equation} 
The ratio $b/k\equiv R_0$ which is recognised as basic reproduction rate, 
quantifies the expected number of secondary
infections from a single infection in a population where all 
individuals are susceptible. Taking derivatives on both sides of the
Equation (\ref{conserv}), the
$i$ equation is obtained as

\begin{equation}
\label{ratei}
\frac{di}{dt}=\left[R_0 \, s(t) \, - 1 \right]i(t).
\end{equation}
At $t=0$, 

\begin{equation}
\label{ratei0}
\frac{di}{dt}|_{t=0}=\left[R_0 \, s(0) \, - 1\right]i(0),
\end{equation}
which shows, for an infection to become epidemic, the condition $(R_0s(0)-1)> 0$ must be satisfied.
Otherwise the infection does not spread but dies out. 

In order to explore the societal/behavioural impact on the pathogen spread, we make $R_0$ time-dependent. 
To fix ideas, notable is the efficacy of a spread depends on the social awareness about the epidemic along with the properties of the pathogen. Such social back-reactions have already been recognized \cite{ferguson}. Funk {\it et.al.} \cite{funk}
have developed a mathematical model which studies the dynamics of an epidemic in the presence of social awareness through either direct observations or rumour. The results document the epidemic dynamics to 
complement human
behaviour and vice versa. Arguably, social awareness can lead to a proactive observance of hygiene---like 
regular hand washing, avoidance of physical contact, and maintaining social distancing. Importantly, the awareness is progressive, {\it i.e.} increases
with time as the epidemic unfolds. For example, individuals may not be aware or fail
to recognize the importance of the above preventive measures until the epidemic significantly develops. Also, 
an aggressive campaign by authorities can implement some of the above deterrents effectively. For example, 
Govt. of India campaigned to raise awareness about the COVID19
by setting an information nugget as a default caller tune across all cell phone service providers. The campaign was particularly effective in rural areas where Internet access is rudimentary, but almost everyone has cell phones. Contrarily, 
it is not practically feasible to implement deterrents 100\% effectively in a finite time.
The reason may either be the consequent recession or
resistance of the populace to the changing lifestyle. 
 To model such a response, we consider

\begin{equation}
\label{bexp}
b(t) = b_0 \, \exp{(-{t}/{\tau})}
\end{equation}
where the time constant $\tau$ determines how fast and effectively a population 
can assimilate preventive interventions. Notably, a monotonically decreasing
$b(t)$ such as the above, only takes into account the social back reactions 
which arrests the epidemic. Contrarily, the back reaction can have a negative impact also. 
For example, propagation of rumors and other misinformations can inhibit the progressive 
social awareness, making $b(t)$ non-monotonic---a scenario excluded  in the present analysis.
 The modified SIR equations are

\begin{eqnarray}
\label{sirs}
& &\frac{ds}{dt} = - b_0 \, \exp{(-t/\tau)} \, s(t) \, i(t) ~,\\
& &\frac{dr}{dt} = k \, i(t) ~,\\
& &\frac{di}{dt} = k \, \left[R_0 (t) \, s(t) - 1\right] \, i(t) ~.
\end{eqnarray}
Importantly, a time dependent decaying $R_0$ opens up the possibility of satisfying 

\begin{equation}
\label{r0less}
\left[R_0 (t)s(t)-1\right] < 1
\end{equation}
during evolution, after which the epidemic fizzles out.

Notably, the SIR model does not differentiate between the clinically recovered population and the deceased but considers both as recovered in a sense that they are no-more susceptible or infected. In this present example, 
it is straightforward to separate the fractional mortality ($\check{r}$) from the recovered ($\hat{r}$) one, 
by simply assuming the fatality rate ($m$), based on the virulence strain of pathogen and also existing 
treatment facility for the age distribution of particular demography. However, a co-morbidity 
can significantly raise the fatality rate which is not considered by the model.  For example, a
 recent paper concludes that patients older than 65 years have more than two times higher risk of 
dying from  COVID-19 while a similar risk exists if the patient is male \cite{ref:1}.
Then 
\begin{equation}
\label{eq:removed}
r(t) = m \, \check{r}(t)  +  (1 - m) \, \hat{r}(t).
\end{equation}
One can extract the rate from the existing data across the different system. It is also expected that the gradual understanding would enable us to rationalize the rate in a subsequent epoch.

Another important extension is the inclusion of lockdown phase mimicked by a sudden reduction of effective $b$ 
($b_{in}$) value for a certain time (lockdown time) like a finite square well. That is implemented in the 
model with two sets of continuous and differentiable Sigmoid functions, such as,

\begin{equation}
\label{eq:pot_well}
b_{lockdown}^{model} =  b_{in} + \frac{(b_{start} - b_{in})}{[1 + \exp{(t - t_{start})}]}
                                                   + \frac{(b_{end} - b_{in})}{[1 + \exp{(t_{end} - t)}] }.
\end{equation}.

Here, $b_{start/end}$ and  $b_{in}$ are corresponding values of effective $b$ before (and after) the lockdown and 
during the lockdown respectively. Similarly, $t_{start}$ and $t_{end}$ are the starting point and end point 
(days) of such lockdown.  Above mentioned single stage lockdown for an extended period of days is neither 
feasible or recommended considering the substantial social and economical cost. It is followed with a multi step lockdown or a staggered removals of lockdown by gradual removal of restrictions. These scenario is studied by 
using two or three staged finite square well developed with Sigmoid functions in a phased manner. Examples are,
\begin{eqnarray}  \nonumber
\label{eq:staggered2}
b_{lockdown, staggered[2]}^{model}  =   b_{in}^{(1)}  &+&
\frac{(b_{start} - b_{in}^{(1)} )} {[1 + \exp{(t - t_{start})}]}
         + \frac{(b_{in}^{(2)} - b_{in}^{(1)} )} {[ 1 + \exp{(t_{in}^{(1)}  - t)}]},\\
&+&
\frac{(b_{end} - b_{in}^{(2)} )} {[1 + \exp{(t_{end} - t)}]},
\end{eqnarray}
with the similar notation as before except two different $b$ values ($b_{in}^{(i=1,2)} $) in two stages of lockdown. 
However, one can also express in economical use of variables in terms of lockdown periods ($\Delta_{t}^{(i)} $) in 
multi stage cases. Three stage model is thus expressed as,
\begin{eqnarray}  \nonumber
\label{eq:staggered3}
b_{lockdown, staggered[3]}^{model}  =   b_{in}^{(1)}  &+& \frac{(b_{start}
- b_{in}^{(1)} )}{[1 + \exp{(t - t_{start})}]}  \\ \nonumber
&+& \frac{(b_{in}^{(2)}  - b_{in}^{(1)} )}{[1 +
\exp{(t_{start} + \Delta_{t}^{(1)}  - t)}]} \\ \nonumber
&+& \frac{(b_{in}^{(3)}  - b_{in}^{(2)} )}{[1 + \exp{(t_{start} + \Delta_{t}^{(1)}  +
\Delta_{t}^{(2)}  - t )}]} \\
&+& \frac{(b_{end} - b_{in}^{(3)} )}{[  1 + \exp{(t_{start} + \Delta_{t}^{(1)} 
+ \Delta_{t}^{(2)}  + \Delta_{t}^{(3)}  - t)}]}  ~.
\end{eqnarray}
Finally, we extend SIR equations to allow for anthropogenic migration from one zone to another. Markedly, the zones can either be separated geographical locations
or a hypothetical separation of the same location into two subzones. The model is developed with the continuous-time approach which results in ODEs where the variables are inherently continuous
and rely on Mathematica's accuracy in solving such equations with a finite time step 
$\Delta t$. The outcome is expected to match with the reality only in the
limit $\Delta t \rightarrow 0$.  In contrast, mathematical models with discrete-time
can be used to solve the SIR equations \cite{alen}.
In a similar work, Zakary et al. devised
a discrete-time SIR model that describes the propagation of
a disease in a population of individuals who travel between
multiple regions \cite{zakary}.

To develop the governing differential equations, two zones: zone 1
and zone 2 are defined such that the total number of susceptible, infected and recovered individuals
satisfy
\begin{eqnarray}
\label{zonetot}
& & S=S_1+S_2, \\ 
& & I=I_1 + I_2, \\
& & R=R_1+R_2. 
\end{eqnarray}
Further, 
\begin{eqnarray}
\label{conserv1}
& & S_1+I_1+R_1=N_1, \\
\label{conserv2}
& & S_2+I_2+R_2=N_2,
\end{eqnarray}
where $N_1$ and $N_2$ are the total populations in 
regions 1 and 2 while $N$ is the overall population. We 
further assume the $N_1$ and $N_2$ to be significantly large such 
that reasonable inter-zonal migration does not affect them: 
in other words, the total populations $N_1$ and $N_2$ are assumed to be 
independently constant.
Corresponding fractions are defined as
\begin{equation}
\label{indvidfarction}
 s_i=S_i/N_i, ~~~
 i_i=I_i/N_i, ~~~
 r_i=R_i/N_i,
\end{equation}
where $i=1,2$.
With $\alpha_{21} S_2(t)$ being the number of susceptible individuals migrating from zone 2 to zone 1 and
$\alpha_{12} S_1(t)$ form zone 1 to zone 2 

\begin{eqnarray}
\label{Si}
& & \frac{dS_1}{dt}\propto \left(1-\alpha_{12}\right)s_1(t)+ \alpha_{21} s_2(t),\\
& & \frac{dS_2}{dt}\propto \left(1-\alpha_{21}\right)s_2(t)+ \alpha_{12} s_1(t),
\end{eqnarray}
and
\begin{eqnarray}
\label{Si}
& & \frac{dS_1}{dt}\propto \left(1-\beta_{12}\right)I_1(t)+ \beta_{21} I_2(t),\\
& & \frac{dS_2}{dt}\propto \left(1-\beta_{21}\right)I_2(t)+ \beta_{12} I_1(t),
\end{eqnarray}
finally leading to 
\begin{eqnarray}
\label{s1}
\frac{dS_1}{dt}&=&-b\left[\left(1-\alpha_{12}\right)s_1(t)+ \alpha_{21} s_2(t)\right]
\left[\left(1-\beta_{12}\right)I_1(t)+ \beta_{21} I_2(t)\right].
\end{eqnarray}

Dividing by $N_1$, we get the $s$-equation for the zone 1

\begin{eqnarray}
\label{s1eqn}
\frac{ds_1}{dt}&=&-b \, \left[\left(1-\alpha_{12}\right)s_1(t)+ \alpha_{21} s_2(t)\right]
[\left(1-\beta_{12}\right)i_1(t)+ \beta_{21} \frac{n_2}{n_1} i_2(t)],
\end{eqnarray}
where $n_1=N_1/N$, $n_2=N_2/N$ and $N_1+N_2=N$. 
A similar derivation for $s_2$ gives

\begin{eqnarray}
\label{s2eqn}
\frac{ds_2}{dt}&=&-b \, \left[\left(1-\alpha_{21}\right)s_2(t)+ \alpha_{12} s_1(t)\right]
[\left(1-\beta_{21}\right)i_2(t)+ \beta_{12} \frac{n_1}{n_2} i_2(t)].
\end{eqnarray}
Similarly, the zonal equations for $r$s  are found to be

\begin{eqnarray}
\label{r1eqn}
& &\frac{dr_1}{dt}=k\left[\left(1-\beta_{12} \right)i_1+\beta_{21}\frac{n_2}{n_1}i_2\right]\\
\label{r2eqn}
& &\frac{dr_2}{dt}=k\left[\left(1-\beta_{21} \right)i_2+\beta_{12}\frac{n_1}{n_2}i_1\right]
\end{eqnarray}
To obtain the $i$ equation, we employ the conservation relations (\ref{conserv1}) and (\ref{conserv2}) in
their fractional form i.e
\begin{eqnarray}
\label{fraczone1}
& &\left(s_1+i_1+r_1\right)=1, \\
\label{fraczone2}
& &\left(s_2+i_2+r_2\right)=1,
\end{eqnarray}

\noindent to generate

\begin{equation}
\label{conservdiff}
\left(\frac{ds_i}{dt}+\frac{di_i}{dt}+\frac{dr_i}{dt}\right)=0,
\end{equation}
where $i=1,2$. Using the $s(t)$ and $r(t)$ equations along with the conditions  
(\ref{fraczone1}) and ({\ref{fraczone2})
form, the $i(t)$-equations for the two zones are obtained as

\begin{eqnarray}
\label{i1eqn}
\frac{di_1}{dt}&=&b \, \left[\left(1-\alpha_{12}\right)s_1(t)+ \alpha_{21} s_2(t)\right]
[\left(1-\beta_{12}\right)i_1(t)+ \beta_{21} \frac{n_2}{n_1} i_2(t)]\nonumber\\
&&- \, k[\left(1-\beta_{12} \right)i_1(t)+\beta_{21} \frac{n_2}{n_1} i_2(t)],
\end{eqnarray}
and 
\begin{eqnarray}
\label{i2eqn}
\frac{di_2}{dt}&=&b \, \left[\left(1-\alpha_{21}\right)s_2(t)+ \alpha_{12} s_1(t)\right]
[\left(1-\beta_{21}\right)i_2(t)+ \beta_{12} \frac{n_1}{n_2} i_1(t)]\nonumber\\
&&- k \, [\left(1-\beta_{21} \right)i_2(t)+\beta_{12} \frac{n_1}{n_2} i_1(t)].
\end{eqnarray}
The six ODEs (\ref{s1eqn}),(\ref{s2eqn}), (\ref{i1eqn}), (\ref{i2eqn}), 
(\ref{r1eqn}) and (\ref{r2eqn})  form a closed set for the 
six variables $s_1$, $i_1$, $r_1$, $s_2$, $i_2$, $r_2$. 
Expectedly, in the absence of migrations from zone 1 to 2 and vice versa 
(${\alpha}_{12}=\alpha_{21}=\beta_{12}=\beta_{21}=0$), the above equations reduce to 
the original SIR equations.

\section{Simulations and Results}
\label{sec:simulation}

\begin{figure}[t]
\centering
\includegraphics[scale=0.85]{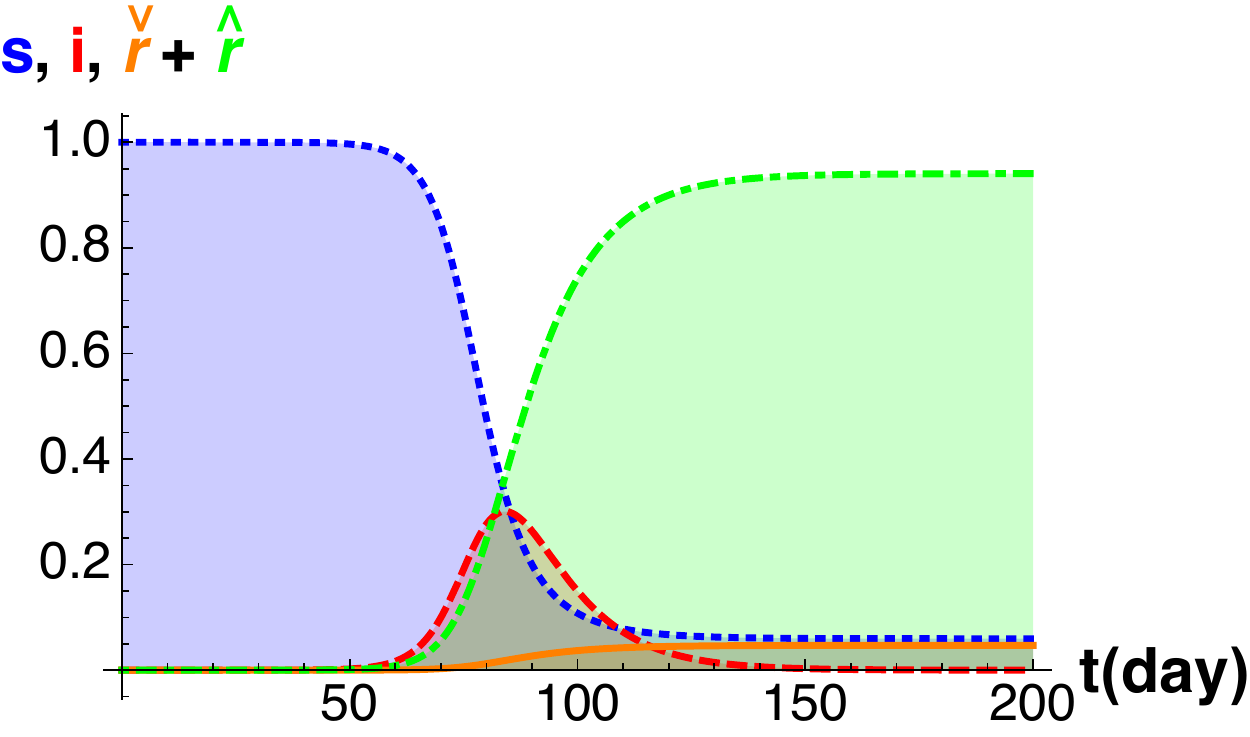}.  
\caption{Demonstration of susceptible rate s(t) in blue, infected rate i(t) in red and removed rate  r(t) in green which includes recovered rate ($\hat{r}(t)$ which is not shown separately) and the mortality rate $\check{r}(t)$ in orange. Parameters considered as 
$b=0.3$, $k=0.1$ and mortality rate $m=0.05$.
}.
\label{sirfig}
\end{figure}

The relevant ODEs are solved by using NDSolve function of the Mathematica with the appropriate initial condition. To benchmark, the following provides results for SIR simulations with initial conditions, 
\begin{eqnarray}
\label{inisir}
& &s\mid_{t=0}=1,\\
& &r\mid_{t=0}=0, \\
& &i\mid_{t=0}= 1\times10^{-7}.
\end{eqnarray}
The initial values are chosen in line with the spread of COVID19 in India. With the 
approximate total population of
India ($\approx 1.00\times{10}^9$) normalized to unity, $100$ infections per day 
yields a normalized value of $i(0)=1\times10^{-7}$, which we use in our calculations. 
Notably, the $100$ infections per day were achieved during the middle of March 2020.
The constant parameter $k$ represent the rate at which the fraction of infected converts into recovered. Assuming an average period of 10 days the pathogen takes to spread the infection, we can choose an approximate $k$ with a fraction of $1/10$.
The solutions are illustrated in the Figure \ref{sirfig} with choice of parameters $b=0.3$, $k=0.1$, amounting to $R_0=3$. 
The histories of $s(t)$, $i(t)$, $r(t)$ are represented by lines of colors 
{\sl{blue, red, green}} respectively. The curve in {\sl{orange}} represents the mortality rate. Notably, the distribution
of $i(t)$ is Gaussian. An increase in $b$ (and hence $R_0$) decreases the Full Width at Half Maxima (FWHM) and the peak of the Gaussian (not shown), indicating a faster spread of the infection. The sum of the mortality and the clinical recovery rates is equal to the recovery rate in standard SIR plots. To distinguish, hereafter, we refer recovery rate in the standard SIR model as the "removed" rate; removed, since the individuals in this category can not be further infected.  The dashed line represents the sum $(s+i+r)$ and is equal to $1$, as expected from the conservation (\ref{conserv}).

\begin{figure}[t]
\centering
 \subfloat[]{\label{sirfigtau_a}\includegraphics[scale=0.55]{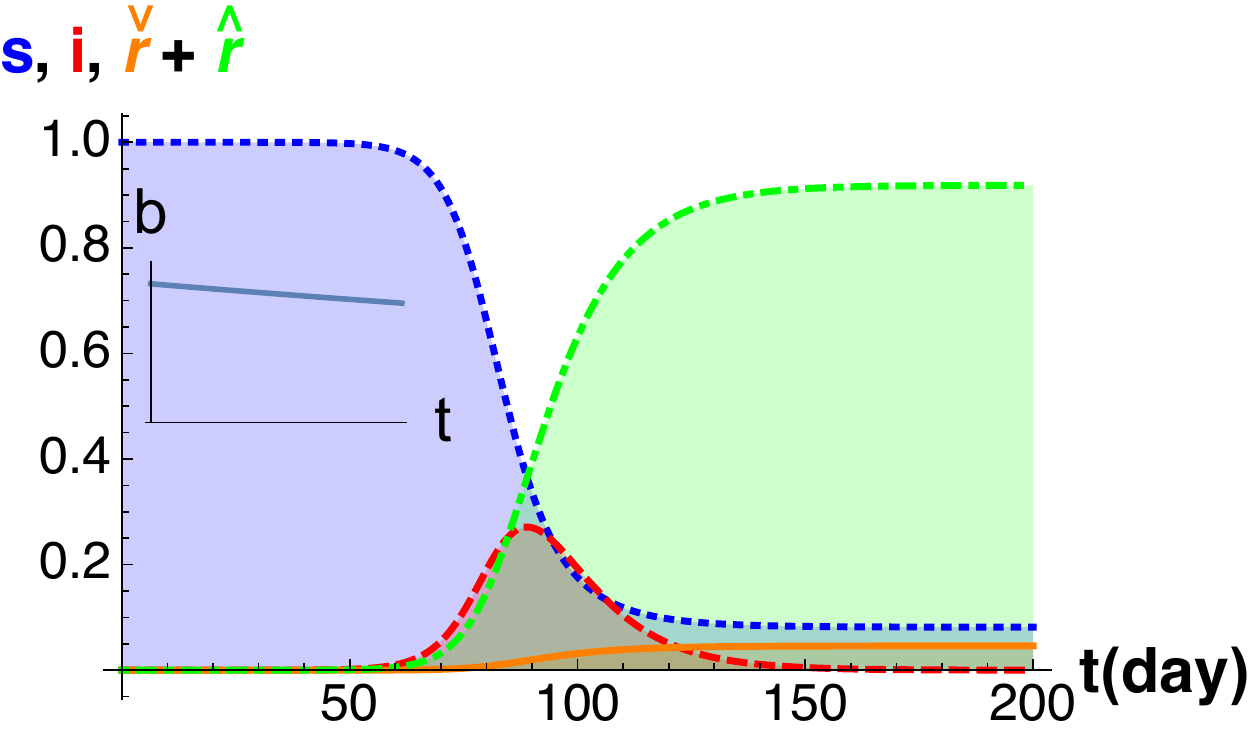}}~ 
 \subfloat[]{\label{sirfigtau_b}\includegraphics[scale=0.55]{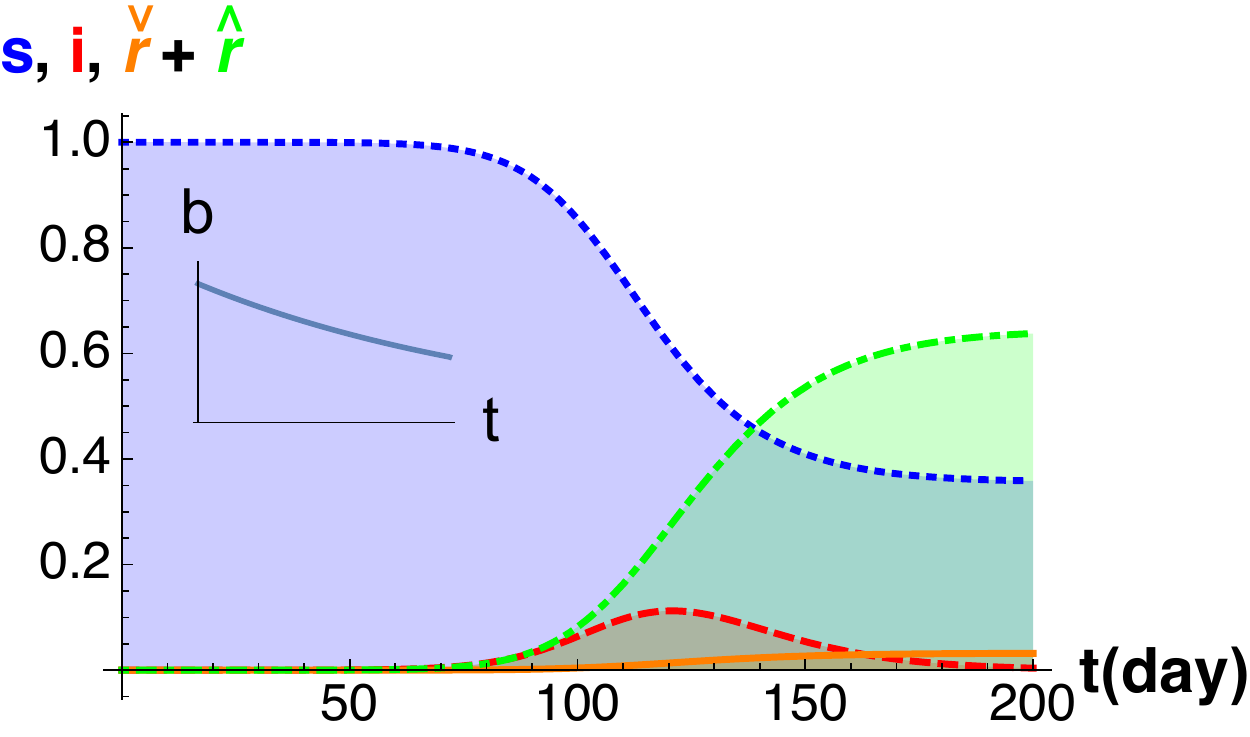}} \\  \vspace{0.4 cm}
 \subfloat[]{\label{sirfigtau_c}\includegraphics[scale=0.55]{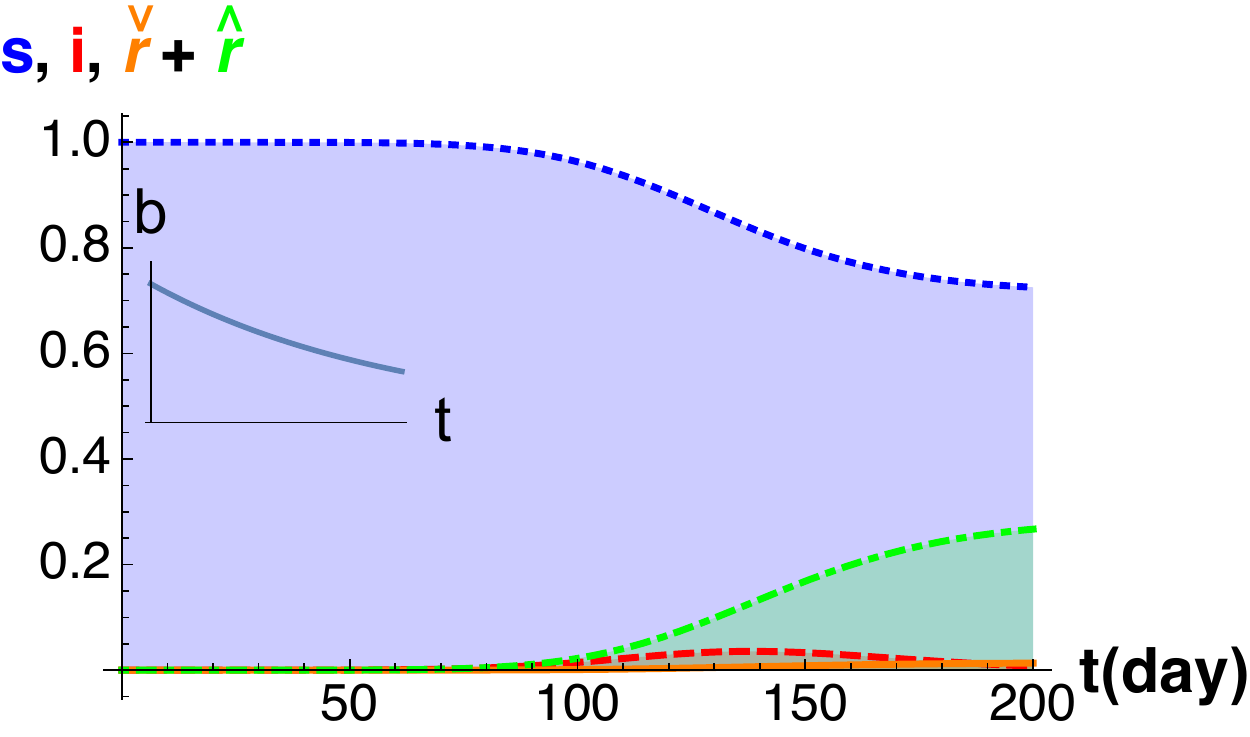}}~ \hspace{0.4 cm}
 \subfloat[]{\label{sirfigtau_d}\includegraphics[scale=0.38]{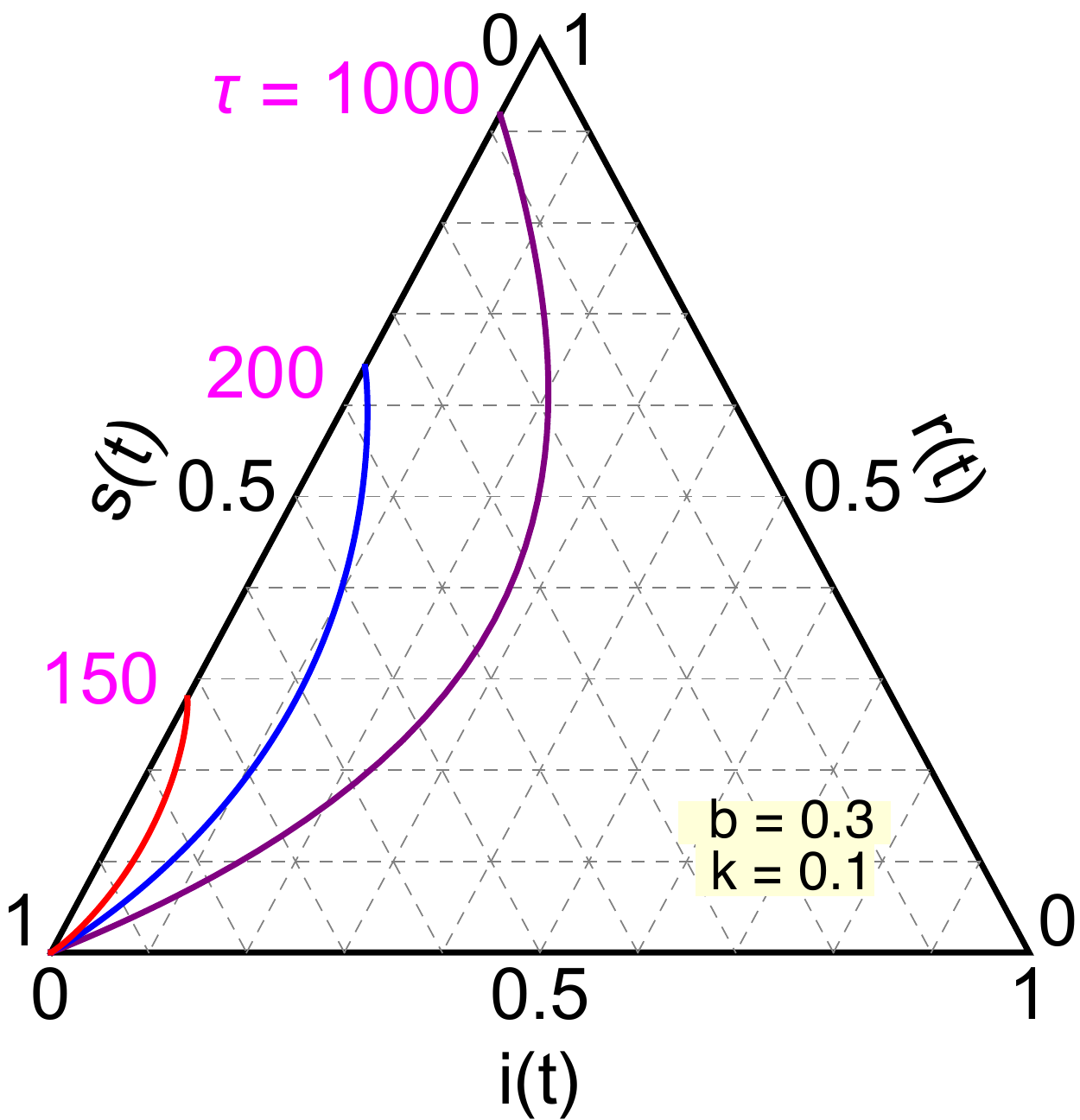}}
\caption{Demonstration of the basic model is infused with progressive social awareness by $\tau =$ (a) 1000 days, (b) 200 days and (c) 150 days respectively. Color scheme, notations and rest of the parameters remains same as Figure \ref{sirfig}. At inset of each plot corresponding variations of $b$ is indicated. In plot (d)  all three plots are described in a ternary diagram.
}
\label{sirfigtau}
\end{figure}

Figure  \ref{sirfigtau} 
represent plots where the basic model is infused 
with the progressive social awareness characterized by $b(t)=b_0 \exp{(t/\tau)}$; the time constant $\tau$ signifies the 
rate at which the $b(t)$ and hence $R_0(t)$ falls, quantifying how fast the society adapts various 
interventions. The inset diagram in \ref{sirfigtau}(a) plots the evolution of $b(t)$ with $b_0=0.3$ and $\tau=1000$ days, which is near constant for all
practical purpose, indicating very little social awareness (or zero intervention) with time. The $s$, $i$ and $r$ plots are, expectedly, identical to the constant $b$ case. The influence of the progressive social awareness is 
evident in the next two plots \ref{sirfigtau}(b) and \ref{sirfigtau}(c) in the same Figure where aggressive 
interventions are imposed with $\tau = 200$ and $150$ days respectively. The insets for both the plots, again, 
show the time variations of $b$. Evidently, the one with the fastest decay as in Figure \ref{sirfigtau}(c) 
exhibits the infection curve to be most flattened and having much lowest peak value. The above findings 
qualitatively agree with the recent simulations by \cite{ref:2} which show delayed onset of 
successively diminished peaks in the total infected population with a stricter adherence to ``social 
distancing".

To further explore the influence of social awareness on infection fraction, 
Figure \ref{sirfigtau}(d)  presents a ternary diagram of the variables $s(t)$, $i(t)$ and $r(t)$
in the $(s, i, r)$ space for different $\tau$ values to quantify their interrelationship. 
In the plot, the time is implicit and $(s, i,r)\in\{0,1\}$ satisfying $s(t)+ i(t)+r(t)=1$ at all $t$. 
To elucidate further, we consider any of the one curve in the plot
and note that at the initial point $s(t=0)=1$, $i(t=0)=1 \times 10^{-7}$, $r(t=0) =0$ representing lower-left corner at the plot. All the three
variables evolve implicitly with time, and after a sufficiently large time interval, all the curves terminate at $i=0$. The  
curve having the largest time constant, 
$\tau=1000$ days (minimally progressing social awareness),  is the highest peaked---having the largest FWHM. The opposite is true for the smallest $\tau=150$ days curve. Notably, 
the three curves with  $\tau=150, 200,1000$ days have three different peak values. 
Importantly, the spacing between the termination points is more for the lower $\tau$ values.
 Contextual to the paper, 
such non-linear dependency implies that a society capable of developing social awareness
at a moderately faster pace during an epidemic gets far more benefited by additional campaigns than the one where the awareness develops at a slower rate. We believe, incorporation of the basic program on epidemic awareness at school curriculum can better prepare a society for a faster response. 
The program will mostly be beneficial for epidemics like COVID19 where in the absence of vaccination and antiviral drugs; social interventions like social distancing, basic respiratory hygiene/cough etiquette,
appropriate hand washing etc., are the only available deterrents.

\begin{figure}[t]
\centering
 \subfloat[]{\label{sirfiglockdown_a}\includegraphics[scale=0.55]{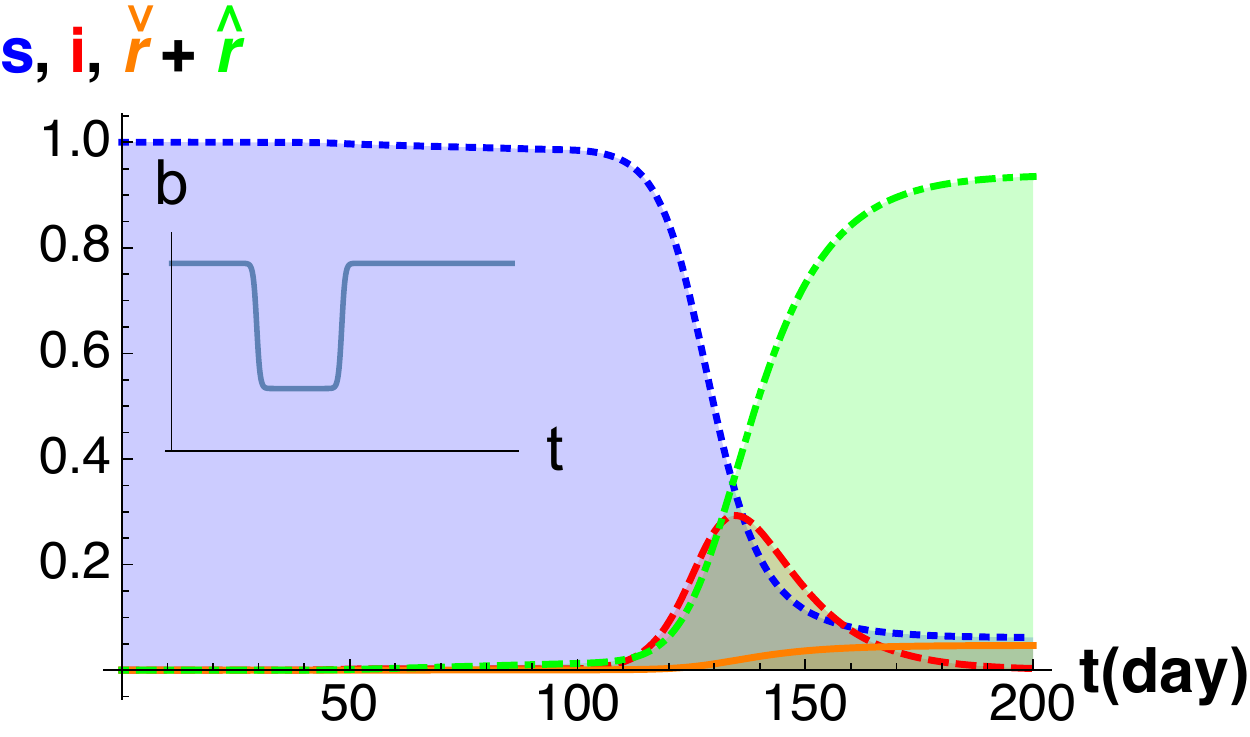}}~ \hspace{0.4 cm}
 \subfloat[]{\label{sirfiglockdown_b}\includegraphics[scale=0.55]{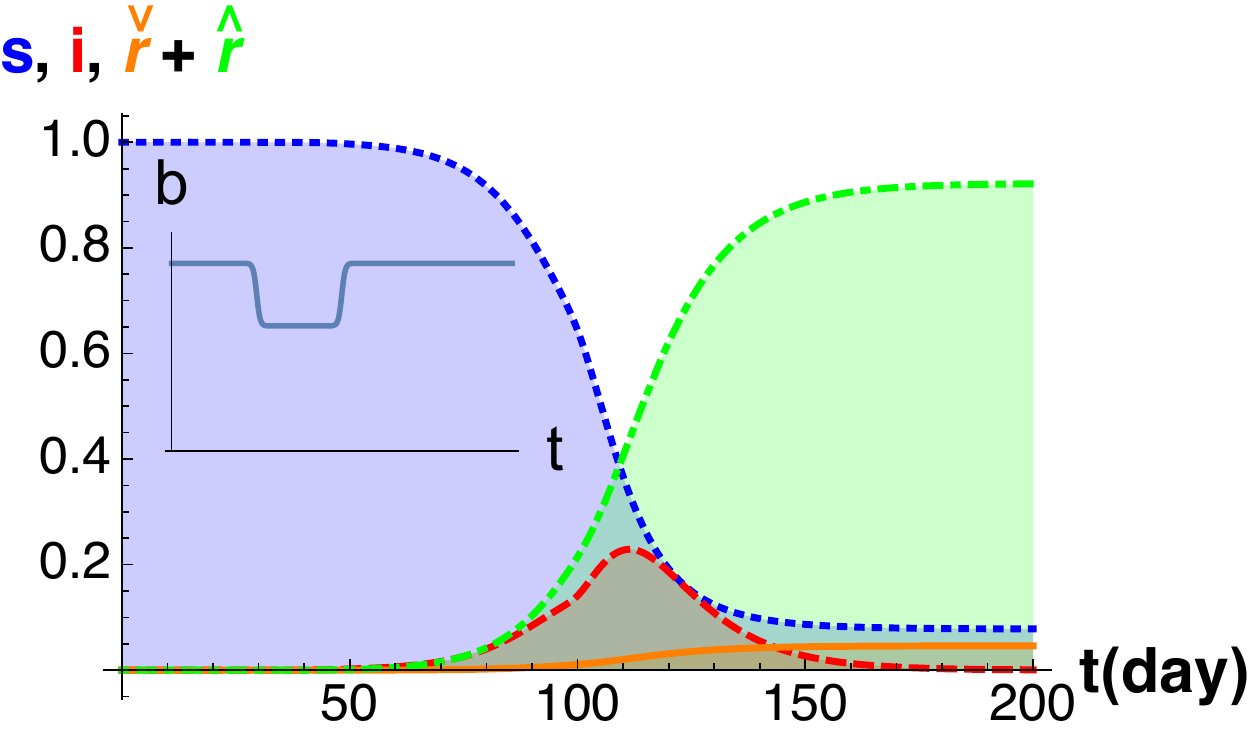}}
\caption{Demonstration of the basic model infused with simplest lockdown mimicked with a $b$ given by a finite square well as shown in the inset of plots with lockdown variables $b_{\rm{in}}=0.1$ and $0.2$. and the  histories of the variables are shown respectively. Color scheme, notations and rest of the parameters remains same as Figure \ref{sirfig}. 
}
\label{sirfiglockdown}
\end{figure}

A complete or a partial lockdown implies an enforced social measure to break the chain of infection by maximizing the social distancing and hence, minimizing the spread. In the following, we discuss the effects of lockdown on the infected fraction without and with progressive social awareness. For the purpose, notable is the realization that the lockdown effectively lowers the value of $R_0$ for some finite period. Hence, the simplest lockdown is mimicked with a sudden reduction of $b$ given by a finite square well. Such condition is modelled with a pair of sigmoid functions as described in Equation \ref{eq:pot_well} and shown as insets in Figure \ref{sirfiglockdown}. Here, in the time evolution of $b$,  the larger value represents no-lockdown, and the smaller value signifies the lockdown.} The plots \ref{sirfiglockdown}(a) and \ref{sirfiglockdown}(b) represent histories of the variables
for $b_{\rm{in}}=0.1$ and $0.2$, respectively while $k=0.1$. 
The no-lockdown value of $b$ in both cases is fixed at $0.3$. The lockdown period ($\Delta_t$) is 50 days, spanning between the day 50 to day 100. Note that effective $R_0^{\rm{in}}$ drops to the values $1$ and $2$ respectively during these example lockdown periods. The choice of low $R_0$ in Figure \ref{sirfiglockdown}(a) effectively stops the infection and idealizes the lockdown to be perfect.
The Figure illustrates two dissimilar peaks in $i(t)$ where the first peak (barely visible for this parameter choice) is in response to the lockdown and is centred at $t\approx 52$. The second peak in $i(t)$ onsets after the lockdown is over and is located at $t\approx 130$, having a value 
$\approx 0.28$ and spread $t\in\{120,180 \}$ days. 
A visual comparison with Figure \ref{sirfig} which documents a similar spread of the 
$i(t)$ curve $t\in\{50,130\}$ and a peak value of $\approx 0.28$ suggests a standalone lockdown with low $R_0^{\rm{in}}$ can 
only delay the peak, providing additional preparation time for the authorities. Second plot in Figure \ref{sirfiglockdown}(b) consider a rather pessimistic lockdown with comparatively higher side of $R_0^{\rm{in}}$ during lockdown. Unlike the previous case, here the trend is more admixture with the no-lockdown scenario, as in Figure \ref{sirfig} except alleviating the curve, reducing the peak and broadening the spread. One realizes that the impact of lockdown period and the choice of $b_{\rm{in}}$ ($R_0^{\rm{in}}$) produces an interplay between double Gaussian in $i(t)$ and their interference which is demonstrated in our next discussion.

\begin{figure}[t]
\centering
 \subfloat[]{\label{sirfiglockdown_contour_a}\includegraphics[scale=0.65]{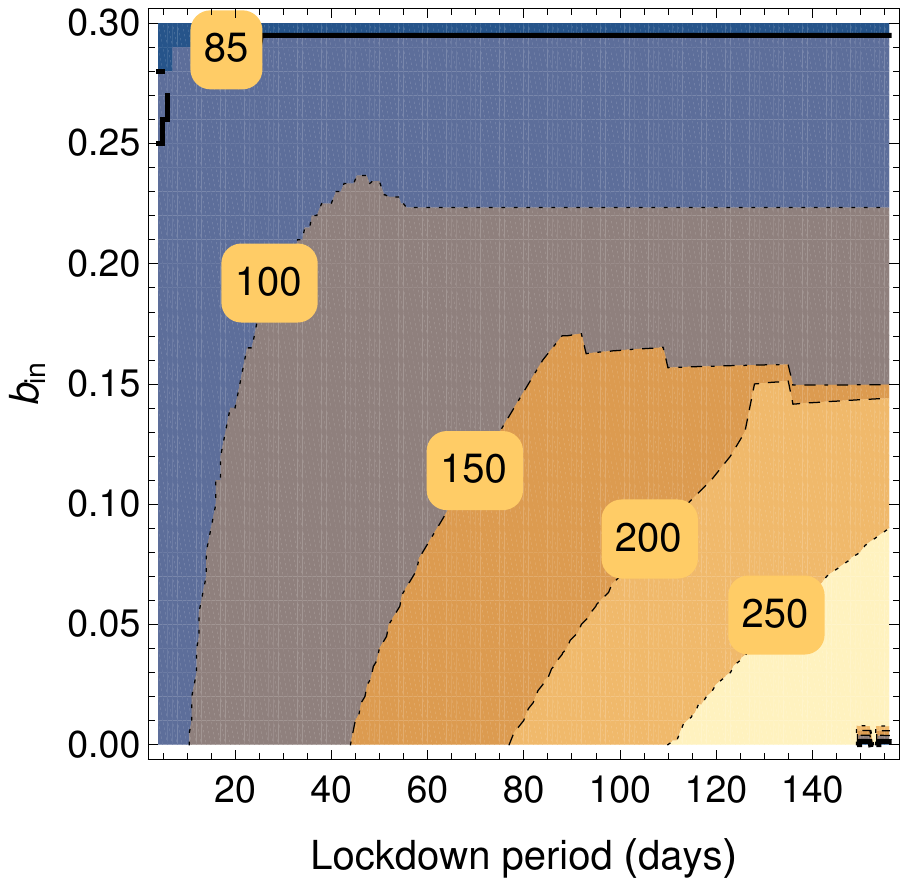}}~ \hspace{0.4 cm}
  \subfloat[]{\label{sirfiglockdown_contour_b}\includegraphics[scale=0.65]{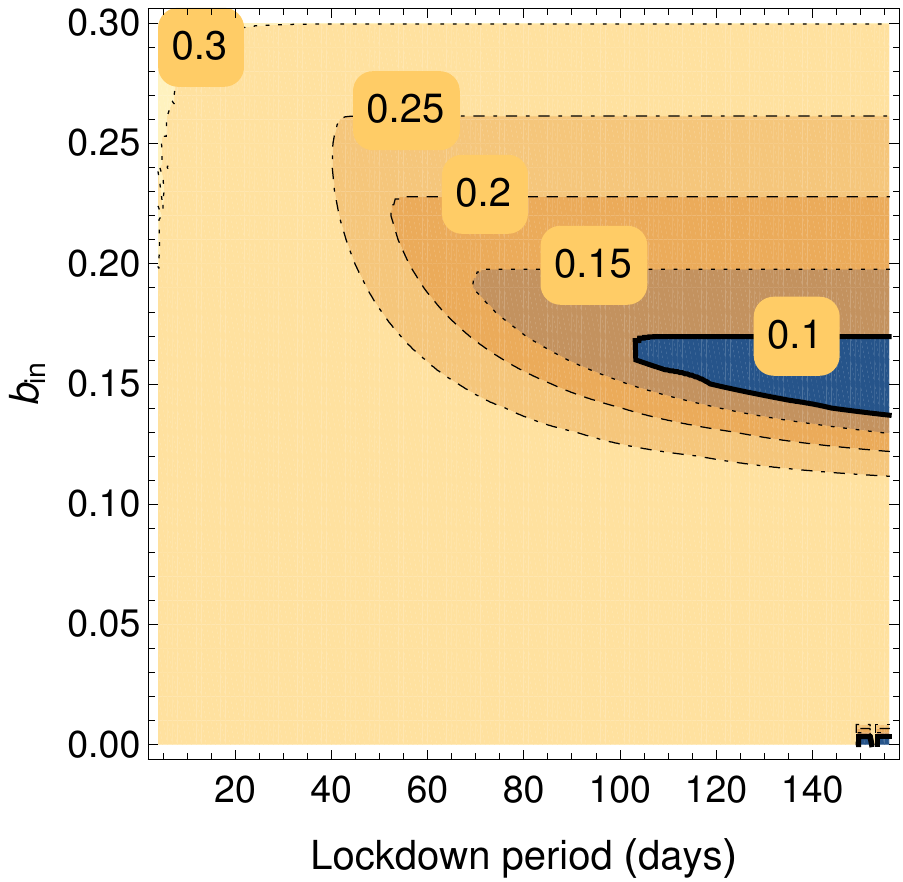}}\\
\caption{Contour plots to demonstrate the (a) day ($t_{i_{max}}$) when infection rate $i(t)$ hits the global maxima and corresponding (b) instantaneous magnitude $i_{max}$  for different choices for lock down period and the  parameter $b_{\rm{in}}$. Rest of the parameters remains same as Figure \ref{sirfig}. 
}
\label{sirfiglockdown_contour}
\end{figure}

The Figures \ref{sirfiglockdown_contour}(a) and  \ref{sirfiglockdown_contour}(b) depict instances ($t_{i_{max}}$ day) of maximum $i(t)$ 
and its magnitude $i_{max}$ as a function of `lockdown period' and the parameter $b_{\rm{in}}$ in contour plots.
Plot \ref{sirfiglockdown_contour}(a) shows the occurrence of the peak value of $i(t)$  is delayed with increasing lockdown period, as expected for low value of $b_{\rm{in}}$ {\it e.g.} in Figure \ref{sirfiglockdown}(a). However for a large $b_{\rm{in}}$ one gets broader distribution with the peak value remaining mostly adjacent to the starting point of the lockdown, as realized  in Figure \ref{sirfiglockdown}(b). Figure \ref{sirfiglockdown_contour}(b)  demonstrates the fact that the peak values $i_{max}$ mostly  remain same irrespective of lockdown period for a fixed  $b_{\rm{in}}$ . However we encounter the same $i_{max}$  twice staying in constant lockdown period. These two peaks correspond to the transition from one to another of the double Gaussian we discussed before. 
Interestingly, for particular range of $b_{\rm{in}} \in \{ b_\alpha, b_\beta \}$, the peak value is minimum for lockdown perion $\Delta_L > \Delta^*$, giving an optimal range of $b_{\rm{in}}$ where a lockdown can be greatly effective. 
 
\begin{figure}[t]
\centering
  \subfloat[]{\label{stagglock_a}\includegraphics[scale=0.55]{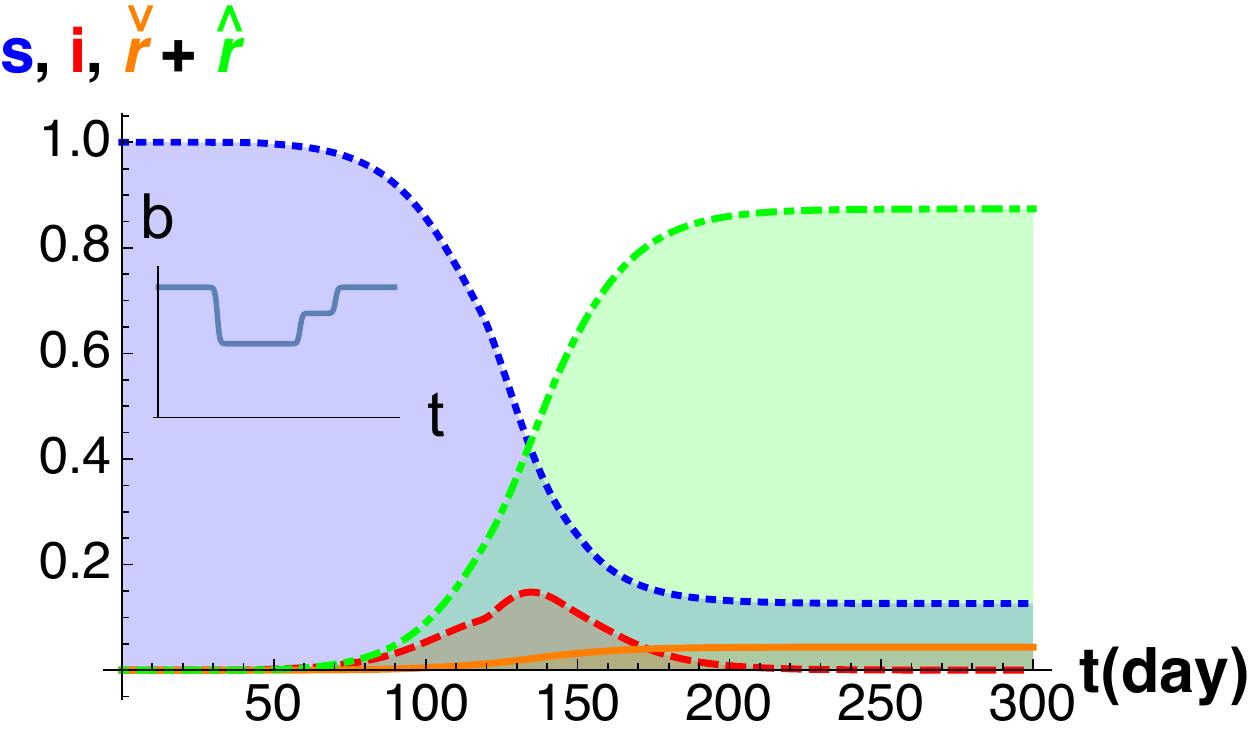}}~ \hspace{0.4 cm}
  \subfloat[]{\label{stagglock_b}\includegraphics[scale=0.38]{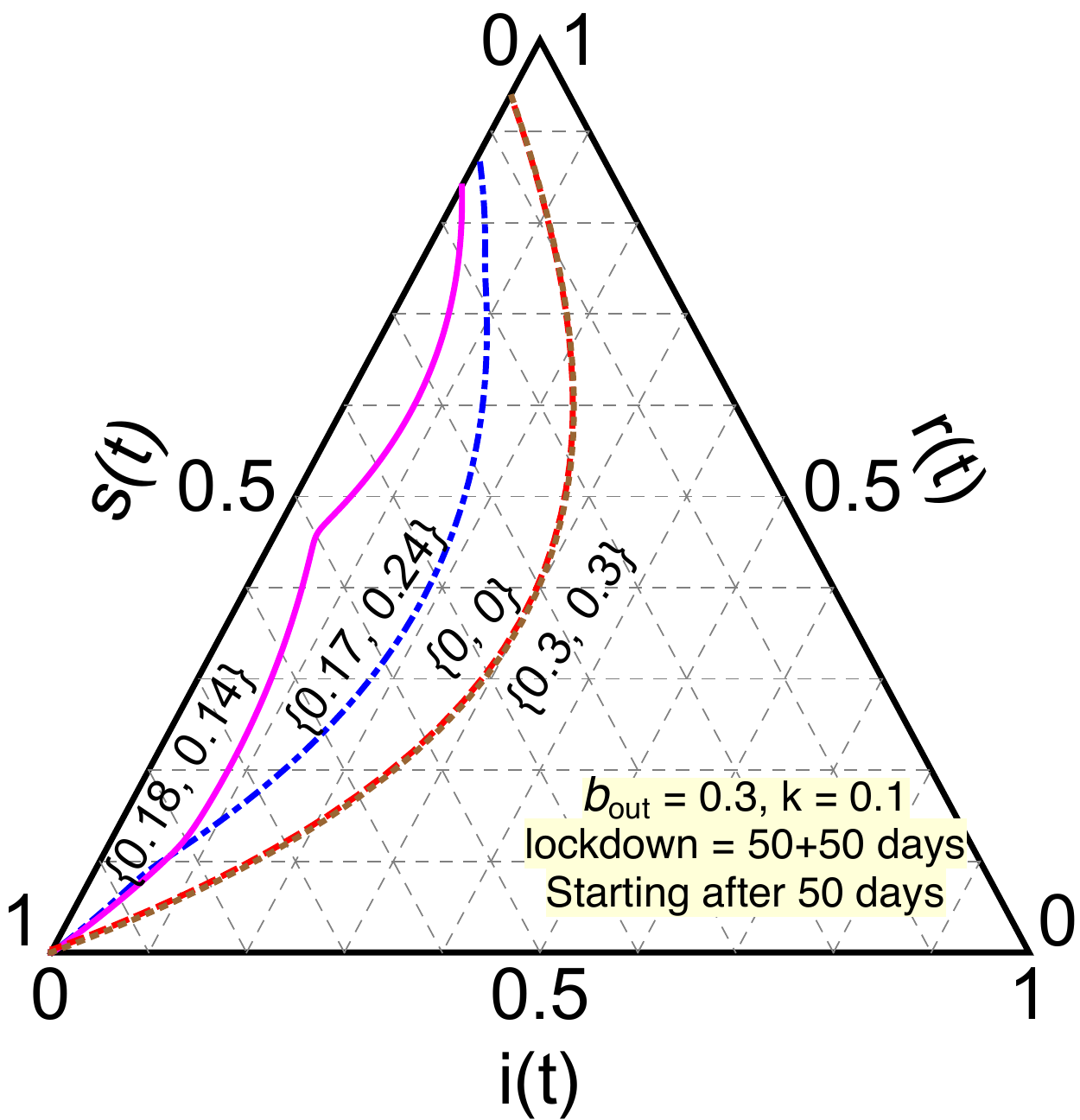}}\\ \vspace{0.4 cm}
   \subfloat[]{\label{stagglock_c}\includegraphics[scale=0.55]{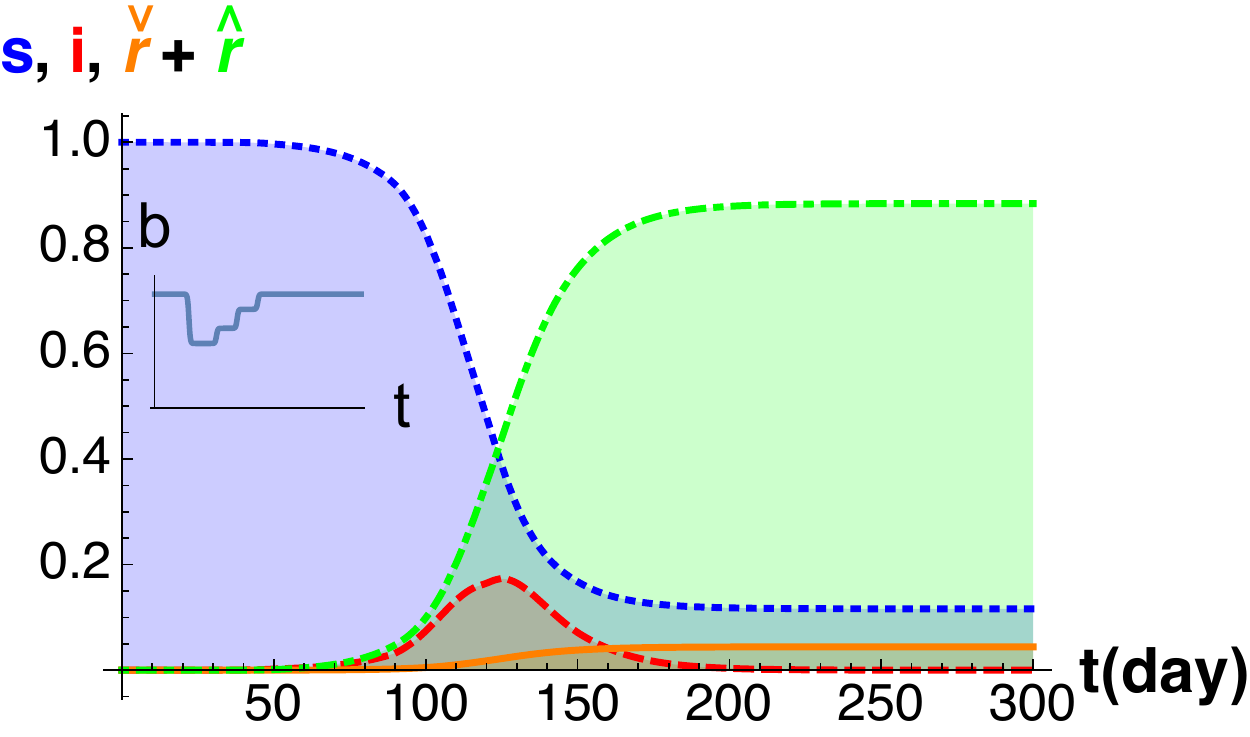}}~ \hspace{0.4 cm}
  \subfloat[]{\label{stagglock_d}\includegraphics[scale=0.38]{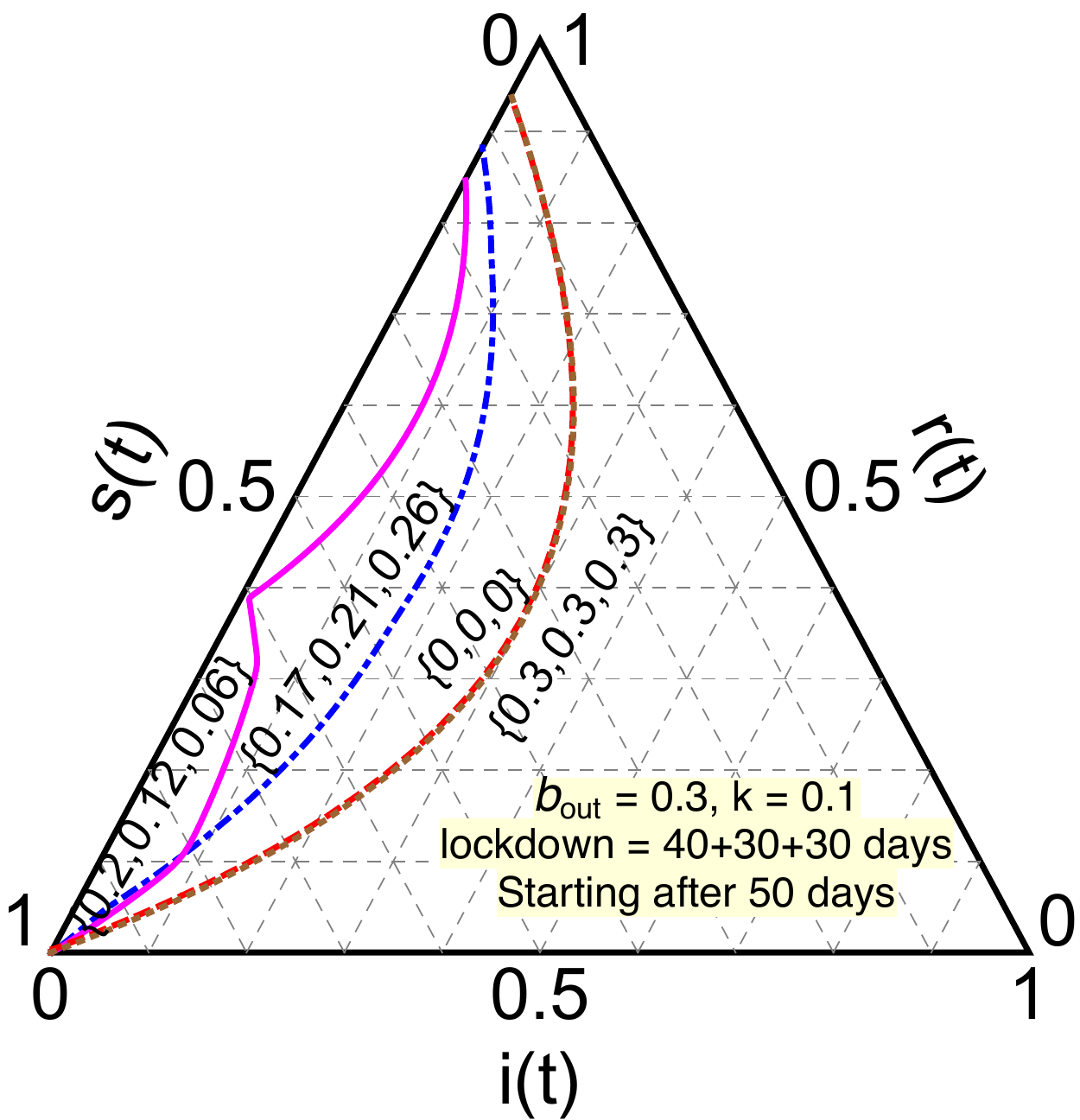}}\\ 
  \caption{Demonstration of the basic model infused with  two step (upper plots)  and three step (upper plots)  staggered removals of lockdown mimicked with a $b$ given by a finite step well as shown in the inset of plots.
Lockdown variables are picked as $b_{\rm{in}}=\{0.17, 0.24\}$  ($\{0.17, 0.21, 0.26\}$) for $\{50 + 50\}$ days  ($\{40 + 30 + 30\}$ days) in two (three) stage staggered removal as demonstrated as inset plots in (a) and (c) respectively . As before, $b$ is always 0.3 outside the lockdown period. Histories of the variables are shown respectively.  In plot (b) and (d) described the running of the variables for the same set of parameters in a ternary diagram along with some other parameters showing for the consistency.}
\label{stagglock}
\end{figure}

\begin{figure}[t]
\centering
  \subfloat[]{\label{tau_stagglock_a}\includegraphics[scale=0.55]{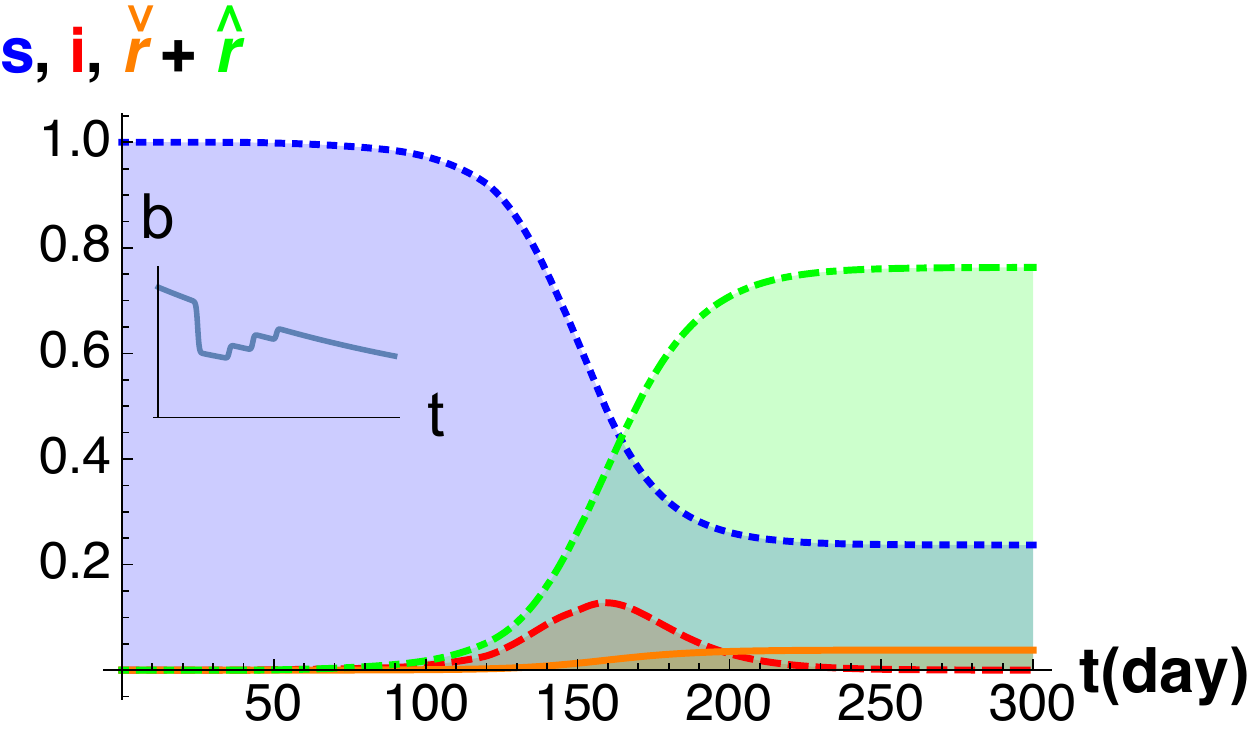}}~ \hspace{0.4 cm}
  \subfloat[]{\label{tau_stagglock_b}\includegraphics[scale=0.55]{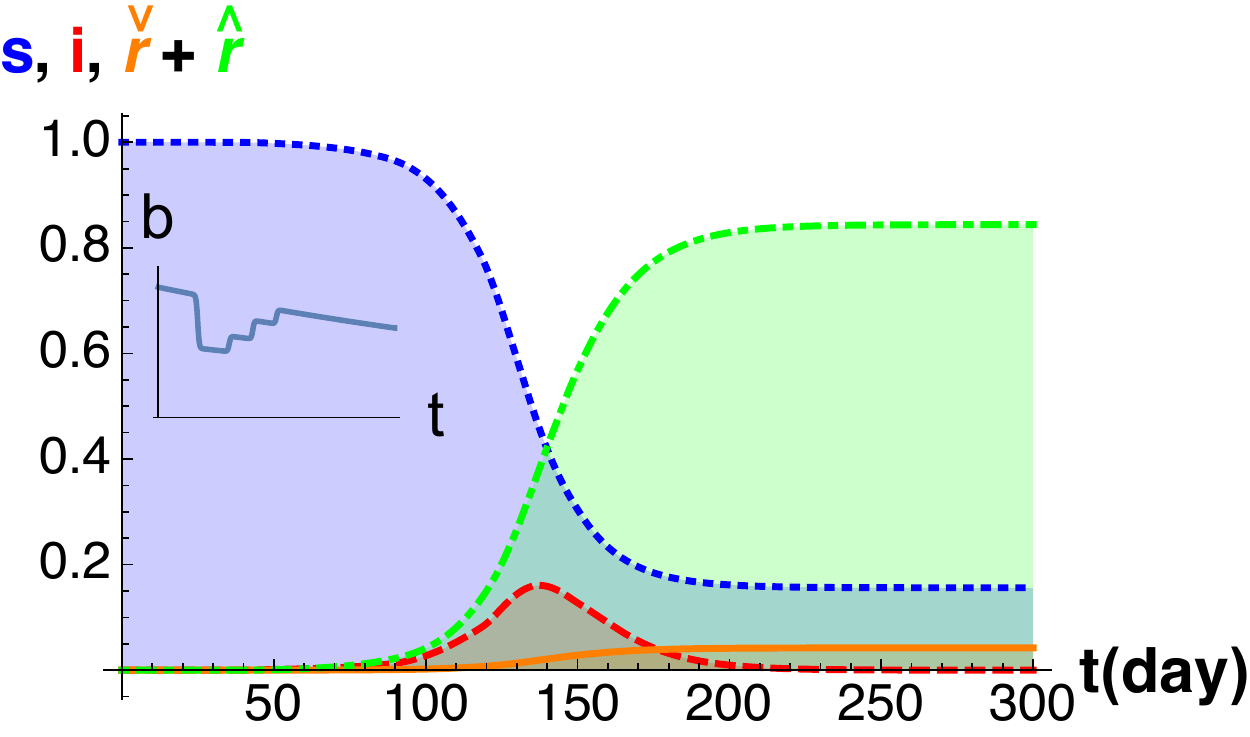}}\\ \vspace{0.4 cm}
  \subfloat[]{\label{tau_stagglock_d}\includegraphics[scale=0.45]{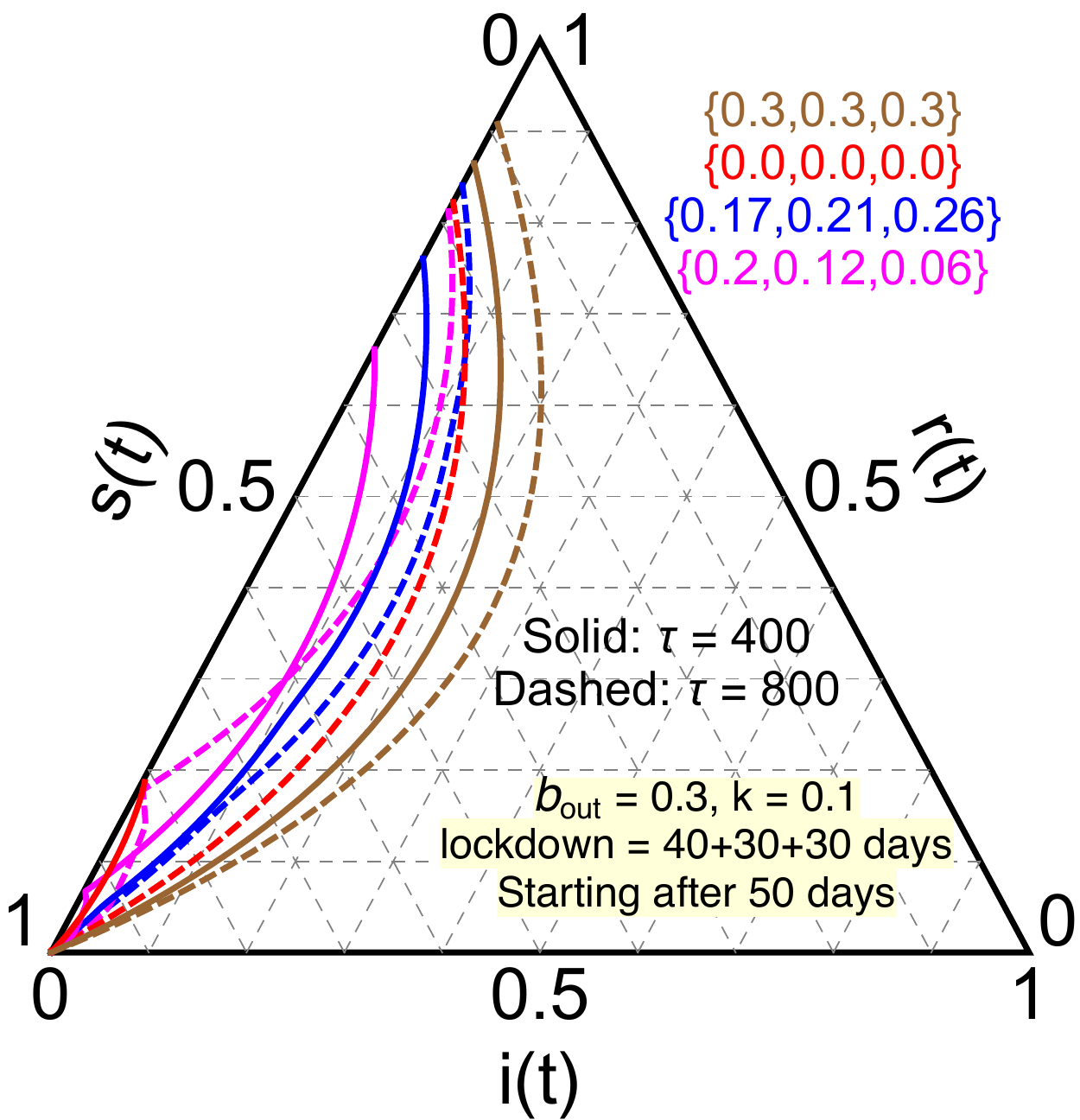}}\\ 
  \caption{Demonstration of the basic model infused with  three-step (upper plots)  staggered removals of lockdown mimicked with a $b$ given by a finite step well as shown in the inset of plots in case of infused with progressive social awareness by $\tau =$ (a) 400 days and (b) 800 days.
Lockdown variables are picked as $b_{\rm{in}} = \{0.17, 0.21, 0.26\}$ for $\{40 + 30 + 30\}$ days in three stage staggered removal as demonstrated as inset plots in (a) and (b) respectively for these two values of $\tau$. As before, $b$ is always 0.3 outside the lockdown period. Histories of the variables are shown respectively.  Plot (c) describe the running of the variables for the same set of parameters in a ternary diagram along with some other parameters showing for the consistency.}
\label{tau_stagglock}
\end{figure}

Further investigations are made to see the effects of staggered removals of lockdown. 
Such removals can be beneficial for the overall economy and also helps the daily wagers to earn their livelihoods together with the fine balance in keeping infection rates manageable.  
Figure \ref{stagglock} illustrates the effects of lockdowns having differently staggered removals. 
Panels \ref{stagglock}(a) and \ref{stagglock}(c) shows the square-step-well functions representing the square-well form of $b$  for a two and three-stage exist after modelling the scenario as described in Equations (\ref{eq:staggered2}) and (\ref{eq:staggered3}).  Characteristic values are 
\{$b_{\rm{in}}^{(1)}=0.17$, $b_{\rm{in}}^{(2)}=0.24$\} and
\{$b_{\rm{in}}^{(1)}=0.17$, $b_{\rm{in}}^{(2)}=0.21$, $b_{\rm{in}}^{(3)}=0.26$\} 
 for two-stage and three-stage staggered removals respectively.
As before, $b$ is fixed at 0.3 outside the lockdown period. The 
$i$-curves in the panels a and b are almost similar, the peak for the
two-staged staggered exit being slightly delayed than the one for the three-staged exit. However, the peak value of the three-stage curve is also somewhat milder than the two-staged one. These characteristics also manifest in the corresponding ternary diagram, where we added a few additional cases for demonstration purpose. Lowest peaked ones (in solid lines) are selected through scanning the parameter space on choosing a set of $b$ values providing minimum peak. One also notices the overlapping lines in perfect lockdown (where $b$'s are zero during lockdown) with ones with no lockdown (where $b$'s don't reduce during lockdown). This demonstrates our earlier argument that perfect lockdown simply delays the curve and show up as overlapping lines in the ternary diagram, where the time axis is implicit in them.

We further add progressive awareness in the above multistage scenarios. Figure \ref{tau_stagglock}
depicts the effect. Once again, the profile of $b$ is in the inset whereas panels (a) and (b)
represent the evolution of the variables for $\tau=400$ and $800$ days, respectively. In both cases, the infection curves are found to be more flattened compared to the one without social awareness. This indicates a staggered exit from the lockdown along with measures to increase social awareness is not only good for an early restart of the economy but also beneficial in flattening the infection curve.

\begin{figure}[t]
\centering
  \subfloat[]{\label{twozone_a}\includegraphics[scale=0.55]{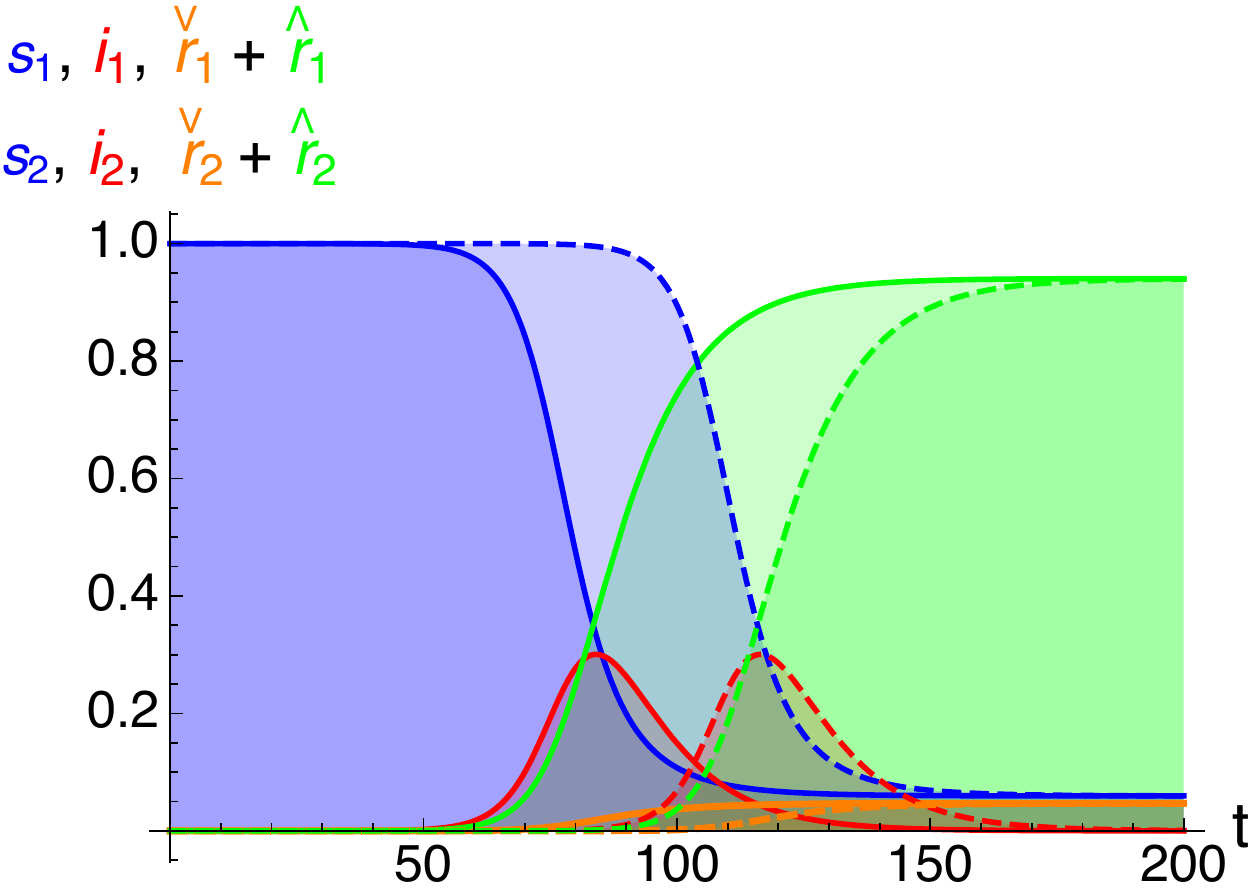}}~
\hspace{0.4 cm}
  \subfloat[]{\label{twozone_b}\includegraphics[scale=0.55]{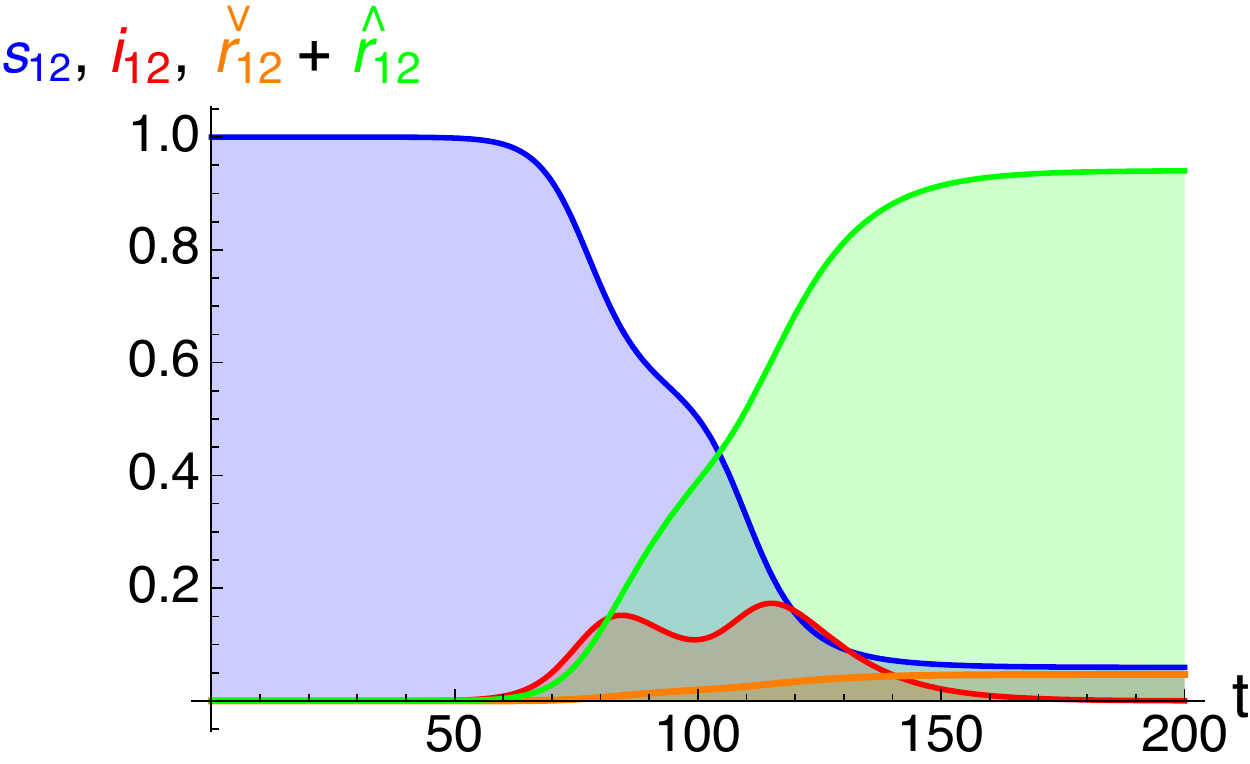}}\\
\
  \caption{Demonstration of the  $SIR$ dynamics in presence of anthropogenic migration 
between two equally populated zones having small inter-zone transfer rates: $\alpha_{12}= \beta_{12} = 10^{-4}$ and $\alpha_{21}= \beta_{21} =10^{-4}$.  Initial small infection rates of $i(t=0) = 10^{-7}$ was introduced only in the first  zone. As expected both infection and recovery started initially in first zone (represented with solid lines). However very soon it was introduced in second zone too, and corresponding rates are shown with dashed lines in left plot.
Combined effect in both zones together is represented in right plot.}
\label{twozone}
\end{figure}

The section is completed with a discussion on the modeling of the effects of migration between
two zones. Notably,
such calculations can easily be extended to include any number of zones. The $s, i, r$ plots for the two zones (the subscripts pointing the specific zone) are depicted in Figure \ref{twozone}.
For simplicity the exchange constants are kept equal and tiny: $\alpha_{12}= \beta_{12} =\alpha_{21}= \beta_{21} =
10^4$, amounting to an equally small number of individuals traveling from zone 1 to zone 2 and
vice versa providing internal mixing among zones.  The initial  $i(t=0)$ in the zones 1 and 2 are selected to be $\{10^{-7}, 0\}$ respectively. 
The histories of the different variables are shown in Figures \ref{twozone_a} and \ref{twozone_b} in solid and dashed curves respectively.
Expectedly, the infection and recovery initially begun at zone 1 and as individuals inter-migrated between
the zones---intensified in zone 2 also (Figure \ref{twozone_a}). The combined evolution is  depicted in 
Figure \ref{twozone_b} which plots $s_{12}=n_1 s_1 +n_2 s_2$, $i_{12}=n_1 i_1 +n_2 i_2$ and $\hat{r}_{12}
+\check{r}_{12}=n_1 \hat{r}_1 +n_1\hat{r}_2+n_2 \check{r}_1+n_2 \check{r}_2$. Future plan involves testing and modeling different combinations to yield targeted favorable outcomes. For example, finding the optimal set of
parameter to effectively contain the spread within a minimal sized zone. which we keep as a future study.

\section{Summary}
\label{sec:summary}
The paper recognizes the importance of social back-reaction 
on the dynamics of an epidemic. In this work, the basic SIR framework is extended to explore the effects of progressive social awareness which is mathematically modelled by a decaying exponential. It is found that the awareness lowers the effective $R_0$ and reduces the peak infection rate while delaying its appearance. Additionally, the progressive awareness is more effective in societies having some seed knowledge about the various social deterrents. Consequently, 
its inclusion in basic school curriculum can be effective in curbing future epidemics like COVID19
where social interventions remain only available deterrents for a significant amount of time. 

The extended model also studies the effects of lockdowns, mimicked by square-well functions generating different effective $R_0$s and having different staggered exit strategies. 
It is found that the simplest lockdown with single-phase implementation and exit
neither flattens the infection curve nor decreases its peak but only delay its appearance.
The additional time can be utilized by the authority to prepare logistics. A staggered exit from a lockdown is better as the strategy flattens the infection curve as well as reduces and
delays the peak. Also, such exist strategies are better from the economical perceptive also.
The most efficient is an exit strategy planned with a joint increase in social awareness. 
Such manoeuvrings can minimize the peak and flattens the infection curve most---which we believe can be beneficial in future epidemics where lockdowns will be necessary. 

We have also extended the SIR model to include two-zone anthropogenic migration.  The example presented is a basic one where an initially equal number of individuals are assumed to populate the two zones. A small number of people are allowed to migrate between the zones
while zone 2 is absolutely infection-free, and zone 1 is characterized with $i(0)=10^{-7}$. With time,
both the regions get substantially infected with $i$ having the same peak and spread while the peak for $i_1$ appearing earlier than $i_2$.

The paper lays a groundwork where different scenarios to arrest the spread of an epidemic along with the importance of social awareness is explored. Although, in the present work, the above scenarios are mostly examined individually; nevertheless, we recognize the natural synergy between progressive social awareness, lockdowns and anthropogenic migration to control the spread of epidemics. Such a study is left as a future exercise.

\section{Acknowledgement}
\label{sec:ack}
The work is supported by Physical Research Laboratory (PRL), Department of Space, Government of India. All the  computations were performed using the HPC resources (Vikram-100 HPC) and TDP project at PRL. 





\begin{thebibliography}{999}

\bibitem{covid19} 
https://www.who.int/emergencies/diseases/novel-coronavirus-2019/technical-guidance/naming-the-coronavirus-disease-(covid-2019)-and-the-virus-that-causes-it

\bibitem{lancet}
lancet Editorial, 
The Lancet doi: https://doi.org/10.1016/S0140-6736(20)30938-7

\bibitem{compmod} 
Sandip Mandal, Ram Rup Sarkar and Somdatta Sinha, 
Malaria Journal 2011, 10:202
http://www.malariajournal.com/content/10/1/202

\bibitem{sir27}
Kermack, W. O. and McKendrick, A. G.,
Proc. Roy. Soc. Lond. A 115, 700-721, 1927.
DOI: https://doi.org/10.1098/rspa.1927.0118

\bibitem{roz} 
{Ganna Rozhnova and  Ana Nunes 2012, Modelling the long-term dynamics of pre-vaccination pertussis, J. R. Soc. Interface. 92959?2970}

\bibitem{web:idmod_model}
https://idmod.org/docs/general/model-si.html

\bibitem{web:idmod}
https://idmod.org/

\bibitem{funk}
Sebastian Funk, Erez Gilad, Chris Watkins, and Vincent A. A. Jansena
Proc Natl Acad Sci U S A. 2009 Apr 21; 106(16): 6872–6877.

\bibitem{ferguson}
 Neil Ferguson, NATURE, {\bf{446}}, 12 April 2007, 733.

\bibitem{ref:1}
Ghosh, Sarada, G. P. Samanta, and Anuj Mubayi. 2020. COVID-19: Regression Approaches of Survival Data in the Presence of Competing Risks: An Application to COVID-19?. Letters in Biomathematics, May.  Available at: https://lettersinbiomath.journals.publicknowledgeproject.org/index.php/lib/ \\ article/view/307

\bibitem{alen} 
Linda J.S. Allen and Amy M. Burgin, Mathematical Biosciences 163 (2000) 1-33

\bibitem{zakary} 
Zakary, O., Rachik, M. and Elmouki, I. 
Int. J. Dynam. Control 5, 917930 (2017). 

\bibitem{ref:2}
Das, Meghadri and Samanta, G.P., A Fractional Order COVID-19 Epidemic Transmission Model: Stability Analysis and Optimal Control (June 5, 2020). Available at SSRN: https://ssrn.com/abstract=3635938
 
\end{thebibliography}
\end{document}